\documentstyle[prl,aps,epsf,twocolumn]{revtex}   
\begin{document}

\title{The Chirality of Exceptional Points } 
  
\author{ W.D.~Heiss$^{1,2}$ and H.L.~Harney$^2$ }

\address{$^1$Department of Physics,
 University of the Witwatersrand,
PO Wits 2050, Johannesburg, South Africa,\\
$^2$Max-Planck-Institut f\"ur Kernphysik, 69029 Heidelberg, Germany}
 
\maketitle    

\begin{abstract} 
Exceptional points are singularities of the spectrum and wave functions which
occur in connection with level repulsion. They are accessible in experiments
using dissipative systems. It is shown that the wave function at an exceptional
point is one specific superposition of two wave functions which are 
themselves specified by the exceptional point. The phase relation of this
superposition brings about a chirality which should be detectable in an 
experiment.
\end{abstract}

\medskip


\narrowtext

Level repulsion is a well known pattern in virtually all aspects of quantum
mechanics. It states that the levels of a selfadjoint Hamiltonian $H$
generically do not cross as a function of a parameter $\lambda $ on which
$H(\lambda )$ depends \cite{wig}. Its 
importance is particularly pronounced in the realm of quantum
chaos \cite{ber,boh}. The connection between exceptional points 
\cite{kato} and the occurrence of level repulsion has been discussed in
\cite{hesa}.

An exceptional point (EP) is a value $\lambda _c$ of the parameter
$\lambda $, where two
of the eigenvalues $E_k$ of $H$ are equal to each other -- say
$E_{\nu }(\lambda _c)=E_{\nu +1}(\lambda _c)$ -- but where the space of the
corresponding eigenvectors is only one-dimensional. We call this a coalescence 
of the eigenvalues and the eigenfunctions $|\psi _{\nu }\rangle,
|\psi _{\nu +1 }\rangle$.
It is well known that this cannot occur for a selfadjoint Hamiltonian, where
$E_{\nu }=E_{\nu +1}$ entails a two-dimensional space of eigenvectors, in which
case the phenomenon is called a degeneracy.

Consider
\begin{equation}
H=H_0+\lambda H_1, \label{oham}
\end{equation}
where $H_0,H_1$ are real and symmetric $N\times N$ matrices, 
and let $\lambda $ be a complex number. Then $H$ is a complex symmetric matrix.
At an EP there is always a singularity -- namely a branch
point -- in the spectrum $E_k(\lambda )$ and the eigenfunctions 
$|\psi _k(\lambda )\rangle$. The spectrum consists of the values that one
analytic function assumes on $N$ Riemannian sheets. The sheets are connected
by $N(N-1)$ square root branch points, the EP's.
If an EP -- connecting $E_{\nu }$ and $E_{\nu +1}$ -- occurs 
 sufficiently close to the real 
$\lambda $-axis, the two levels undergo a level repulsion as $\lambda $ 
sweeps over the real axis in the vicinity of the EP. Conversely, when two 
levels undergo repulsion, there is always a nearby EP, where
the expansions
\begin{eqnarray} 
E_{\nu }(\lambda )
&=&E^0_{\nu }+\sum_{s=1}^{\infty} e_s (\sqrt{\lambda -\lambda _c})^s  \cr
E_{\nu +1}(\lambda )
&=&E^0_{\nu }+\sum_{s=1}^{\infty} e_s (-\sqrt{\lambda -\lambda _c})^s  
\end{eqnarray}
exist with a finite radius of convergence. 

Three major results have  been shown in \cite{he99} and experimentally 
verified in \cite{demb} when an EP is encircled in the complex 
$\lambda $-plane:
\begin{enumerate}
\item  The two energy
levels $E_{\nu }$ and $E_{\nu +1}$ connected at the EP are interchanged
by a complete turn in the $\lambda $-plane. 
\item The two wave functions $|\psi _{\nu }\rangle $ and 
$|\psi _{\nu +1}\rangle $ are not just 
interchanged like their eigenenergies but one of them undergoes a 
change of sign. In other words, a complete loop in the $\lambda $-plane leads
to $\{\psi _{\nu },\psi _{\nu +1}\}\to \{-\psi _{\nu +1},\psi _{\nu }\}$. 
As an immediate 
consequence we conclude: (i) the EP is a fourth order branch point for the 
wave functions and (ii) different directions of going through the loop yield 
different phase behavior. In fact, encircling the
EP a second time in the same direction we obtain
$\{-\psi _{\nu },-\psi _{\nu +1}\}$ while the next loop yields
$\{\psi _{\nu +1},-\psi _{\nu }\}$ and only the fourth loop restores the
original pair $\{\psi _{\nu },\psi _{\nu +1}\}$. It follows that the opposite 
direction yields after the first loop what is obtained after three
loops in the former case.
\item The behavior of the two energy levels is distinctly different when
a path in the $\lambda $-plane is taken below or above an EP. In one of the
cases, the two levels avoid each other while their widths cross, in the
other case, the two levels cross while their widths avoid each other.
\end{enumerate}

In \cite{demb} the topological structure of an EP has  
been shown in the laboratory to be a physical reality. 
In the present paper, we focus attention 
upon the wave function at the EP. We show that the chiral behaviour that
appears under item (2) above, is an intrinsic property of an EP.  
We argue that this chiral behaviour should be detectable in a suitable
experiment. 

Recall that for $\lambda \to \lambda _c$, one
has $|\psi _{\nu }(\lambda )\rangle \to |\psi_{{\rm EP}}\rangle $ and 
$|\psi _{\nu +1}(\lambda )\rangle
\to |\psi_{{\rm EP}}\rangle $ for the two coalescing wave functions.
We mention that all the other $N-2$ wave functions are regular at a given EP. 

Since $H$ of Eq.(\ref{oham}) is not selfadjoint for complex $\lambda $, the
right hand eigenvectors $|\psi _k\rangle $ are different from the left hand
eigenvectors $\langle \tilde \psi _k|$. Both systems together form a
biorthogonal basis, i.e.~the completeness relation reads for 
$\lambda \ne \lambda _c$
\begin{equation}
\sum _k {|\psi _k\rangle \langle \tilde \psi _k|\over \langle \tilde \psi _k|
\psi _k\rangle }=1. \label{comp}
\end{equation}
Recall that
\begin{equation}
 \langle \tilde \psi _j|\psi _{j'}\rangle =0, \quad j\ne j'. \label{orth}
\end{equation}
Due to the symmetric form of $H$
the left hand eigenvector $\langle \tilde \psi |$ is just the complex 
conjugate of its right hand partner. Hence, in the Dirac notation, the
(complex) components of the row vector $\langle \tilde \psi |$ coincide with
the components of the column vector $| \psi \rangle $. 
From Eq.(\ref{orth}) it follows that
\begin{equation}
\langle \tilde \psi _{{\rm EP}}|\psi _{{\rm EP}}\rangle =0 \label{zero}
\end{equation}
 since the
orthogonality holds identically in $\lambda $ and thus in particular at
$\lambda =\lambda _c$, when $j=\nu $ and $j'=\nu +1$. As a consequence, 
the inverse of the biorthogonal norm $ \langle \tilde \psi _k|
\psi _k\rangle $ that appears in Eq.(\ref{comp}), does not exist at
$\lambda = \lambda _c$ for $k=\nu ,\nu +1$.

For a two-dimensional space, $N=2$, one concludes from Eq.(\ref{zero})
that $|\psi _{{\rm EP}}\rangle $ has the form
\begin{equation}
|\psi _{{\rm EP}}\rangle \sim \pmatrix{\pm i \cr 1 }. \label{gen}
\end{equation}
Nowhere in the foregoing, a basis has been laid down with respect to which
the coefficients of the vector are to be taken. In fact, (\ref{gen})
remains true under all orthogonal transformations of a given basis, even 
complex orthogonal ones. These are the transformations that conserve the 
symmetry of $H$ which we consider. Hence, in every basis with
respect to which $H$ is symmetric, $|\psi _{{\rm EP}}\rangle $ will have 
the form (\ref{gen}). In particular, there is no orthogonal transformation
that maps $\pmatrix { i \cr 1}$ onto $\pmatrix {-i \cr 1}$. Every 
$|\psi _{{\rm EP}}\rangle $  is 
therefore either $\sim \pmatrix { i \cr 1}$ or $\sim \pmatrix {-i \cr 1}$. 

For illustrative purpose we consider a two level model in detail. We 
stress, however, and below explicitly elaborate that even an infinite
dimensional problem is, in the vicinity of an EP, locally equivalent to a two
dimensional problem. 

Consider
\begin{equation}  \label{ham}
H=\pmatrix {\epsilon _1 & 0 \cr 0 & \epsilon _2}+
\lambda U \pmatrix {\omega _1 & 0 \cr 0 & \omega _2} U^T
\end{equation}
with
\begin{equation} \label{uang}
U(\phi )=\pmatrix {\cos \phi & -\sin \phi \cr \sin \phi & \cos \phi },
\end{equation}
where the angle $\phi $ and the energies $\epsilon _k, \omega _k,\, k=1,2$
are real.
The eigenvalues are
\begin{equation}  \label{eigv}
E_{1,2}(\lambda )={\epsilon _1+\epsilon _2+\lambda (\omega _1+\omega _2)
\over 2} \pm R,
\end {equation}
where
\begin{eqnarray} \label{res}
R=&\biggl\{&({\epsilon _1-\epsilon _2\over 2})^2 
+({\lambda (\omega _1-\omega _2)\over 2})^2  \\
&+&{1\over 2}
\lambda (\epsilon _1-\epsilon _2)(\omega _1-\omega _2)\cos 2 \phi
\biggr\}^{1/2}. \nonumber
\end{eqnarray}
The two levels coalesce when $R(\lambda )$ vanishes. This happens at
\begin{equation} \label{exc}
\lambda _c^{\pm }=-{\epsilon _1-\epsilon _2\over \omega _1-\omega _2}
\exp (\pm 2i\phi ).
\end{equation}
Note that for zero coupling ($\phi =0$) the two branch points cancel each 
other and a genuine degeneracy occurs with the well known properties of a 
diabolic point \cite{ber}.

The eigenfunctions can be parametrised by the complex angle 
$\theta $ as follows
\begin{equation} \label{psi}
|\psi _1(\lambda )\rangle =\pmatrix{\cos \theta \cr \sin \theta }, \quad
|\psi _2(\lambda )\rangle =\pmatrix{-\sin \theta \cr \cos \theta }.
\end{equation}
Here, $\theta $ is related to the parameters in Eq.(\ref{ham}) {\it via}
\begin{eqnarray} \label{tan}
\tan \theta (\lambda )&=&   
\lambda ( \omega _1-\omega _2)\sin 2 \phi /  \\ \nonumber
(E_1(\lambda )  &-&E_2(\lambda )+\epsilon _1-\epsilon _2+  
\lambda (\omega _1-\omega _2)\cos 2 \phi ).
\end{eqnarray}
The eigenvectors of Eq.(\ref{psi}) are normalised in the biorthogonal sense
\begin{equation}
\langle \tilde \psi _k|\psi _k\rangle =1 \quad {\rm for} \quad k=1,2,\,
\lambda \ne \lambda _c.
\end{equation}
An explicit calculation shows that the coefficients of the wave functions  
Eq.(\ref{psi}) diverge at the EP. Inserting $\lambda _c^{\pm }$ into
Eq.(\ref{tan}) we obtain
$$ \tan \theta _c^{\pm }=\mp i .$$
This implies for $\lambda \to \lambda _c^{\pm }$
\begin{eqnarray} 
\cos \theta ^{\pm }&\to & \infty \cr
\sin \theta ^{\pm }&\to & \mp i \infty.
\end{eqnarray}
While the completeness relation Eq.(\ref{comp}) is obeyed for those values of 
$\lambda $ which do not coincide with an EP, the set of eigenfunctions is 
incomplete at an EP. At the EP, the eigenfunctions (\ref{psi}) coalesce 
for $\lambda \to \lambda _c^{\pm }$ as
\begin{eqnarray}
|\psi _1(\lambda )\rangle &\to &F_1 \pmatrix {\pm i \cr 1}, \nonumber \\
|\psi _2(\lambda )\rangle &\to &F_2 \pmatrix {\pm i \cr 1}. \label{ep}
\end{eqnarray}
Here, the factors $F_{1,2}$ depend on $\lambda $; in fact they diverge for
$\lambda \to \lambda _c$.

So far we have established that, in the two dimensional model, the wave 
function at the EP has, up to a complex factor, a strictly
prescribed form: the ratio of the components 
is $+i$ at $\lambda _c^+$ and $-i$ at $\lambda _c^-$.
This holds in any basis and {\sl irrespective of
the parameters} $\epsilon _i,\,\omega _i$ and $\phi $. Before we turn to the 
physical relevance of this result we discuss a higher dimensional
situation as this would usually prevail in an experimental set-up.

In higher dimensions the wave function at the EP will of course no
longer have the simple form of Eq.(\ref{ep}), in fact 
$|\psi_{{\rm EP}}\rangle $ has then $N$ components. But using the 
completeness relation
of Eq.(\ref{comp}) we expand
\begin{equation}
|\psi_{{\rm EP}}\rangle =\sum _k c_k(\lambda )|\chi _k(\lambda )\rangle 
\label{expa}
\end{equation}
with
$$c_k=\langle \tilde \chi_k|\psi _{{\rm EP}}\rangle,$$
where
$$|\chi_k\rangle 
={|\psi _k \rangle \over \sqrt{\langle \tilde \psi_k|\psi_k \rangle }}$$
and
$$\langle \tilde \chi_k| 
={\langle \tilde \psi _k| \over \sqrt{\langle \tilde \psi_k|\psi_k \rangle }}$$
which ensures biorthogonal normalisation. If this expansion is used in close
vicinity of an EP, where $|\psi _{\nu }\rangle $ and 
$|\psi _{\nu +1}\rangle $ are about to
coalesce, it is obvious that only the terms
in Eq.(\ref{expa}) with $k=\nu $ and $k=\nu +1$ 
make substantial contributions. 
In fact, all $c_k$ vanish when $\lambda \to \lambda _c$ as follows from
the orthogonality for $k\ne \nu ,\nu +1$ and from Eq.(\ref{zero}) 
for $k=\nu ,\nu +1$.
However, the vanishing numerators for $k=\nu ,\nu +1$ are compensated by the 
vanishing denominators with the result that in the limit
$\lambda \to \lambda _c$ only the terms with $k=\nu $ and $k=\nu +1$ survive. 
This result implies that the $N$-dimensional vector 
$|\psi _{{\rm EP}}\rangle $ is basically
a superposition of only the two ($N$-dimensional) vectors 
$|\psi _{\nu }(\lambda )\rangle $ and $|\psi _{\nu +1}(\lambda )\rangle $; 
the closer $\lambda $ is to $\lambda _c$ the 
more correct is the statement. In other words, with regards to the EP,
the $N$-dimensional problem can be locally simulated by a two-dimensional 
problem. From Eq.(\ref{ep}) we thus conclude that
\begin{equation}
{c_{\nu } \over c_{\nu +1}}=+i \quad {\rm or}
\quad {c_{\nu } \over c_{\nu +1}}=-i \label{pun}
\end{equation}
must hold in the vicinity of $\lambda _c$ but {\sl independent} of $\lambda $
within this vicinity. 

A more explicit analytic 
consideration shows how this astounding and important result comes about.
We denote the components of $|\psi _{{\rm EP}}\rangle $ by 
$\{x_k\},k=1,\ldots,N$ and
recall (Eq.(\ref{zero})) that $\sum _k x_k^2=0$. If $\lambda $ is near 
to $\lambda _c$ the
components of the unnormalised $|\psi _{\nu }(\lambda )\rangle $ can 
be chosen as
$\{x_k+d_k\}$ with 
$d_k= a_k\sqrt{\lambda -\lambda _c}+O(\lambda -\lambda _c),\,k=1,\ldots,N$ and
some constants $a_k$ being of no interest here. The
components of the unnormalised $\langle \tilde \psi _{\nu +1}(\lambda )|$ must 
therefore, to lowest order in the $d_k$, have
the form $\{x_k-d_k\}$. To lowest order in the $d_k$ we obtain
\begin{eqnarray}
\langle \tilde \psi _{\nu }|\psi _{\nu }\rangle &=& 2\sum_k x_kd_k \\ 
\langle \tilde \psi _{\nu +1}|\psi _{\nu +1}\rangle &=& -2\sum_k x_kd_k \\ 
\langle \tilde \psi _{\nu }|\psi _{{\rm EP}}\rangle &=& \sum_k x_kd_k \\ 
\langle \tilde \psi _{\nu +1}|\psi _{{\rm EP}}\rangle &=& -\sum_k x_kd_k 
\end{eqnarray}
from which the statement of Eq.(\ref{pun}) immediately follows. 

The local reduction -- in the vicinity of an EP -- of the 
full $N$-dimensional problem to an effective two dimensional problem is now
achieved by the two-dimensional matrix $h=h_0+\lambda h_1$ with the
matrix elements
\begin{eqnarray}
(h_0)_{jj'}&=&\langle \tilde \chi _j|H_0|\chi _{j'}\rangle \label{h0} \\
(h_1)_{jj'}&=&\langle \tilde \chi _j|H_1|\chi _{j'}\rangle ,\quad j,j'=\nu ,\nu +1 
\label{h1}
\end{eqnarray}
using the {\sl relevant} state vectors $|\chi _{\nu }\rangle $ and 
$|\chi _{\nu +1}\rangle $.
In Fig.1 we display two different but typical examples to 
demonstrate how efficiently the 
procedure works. The eigenvalues of $h_0$ and $h_1$ yield the effective values
of the $\epsilon _j$ and $\omega _j$ as used in Eq.(\ref{ham}). The 
effective coupling angle $\phi $ is obtained from the eigenvectors of $h_1$
in the basis where $h_0$ is diagonal (note that $h_0$ from Eq.(\ref{h0})
is not {\it a priori} diagonal). The straight lines in Fig.1 are
the lines $\epsilon _j+\lambda \omega _j$ which correspond to the effective
unperturbed lines ($\phi =0$); switching on $\phi $ to the calculated value
yields an almost exact approximation of the $N$-dimensional problem 
by the effective two-dimensional problem. 
For each level repulsion, i.e.~for each EP, the procedure has to 
be carried out from the outset. The example of Fig.1 is based on a 
random ten dimensional case.

\begin{figure}
\epsfxsize=2.2in
\centerline{
\epsffile{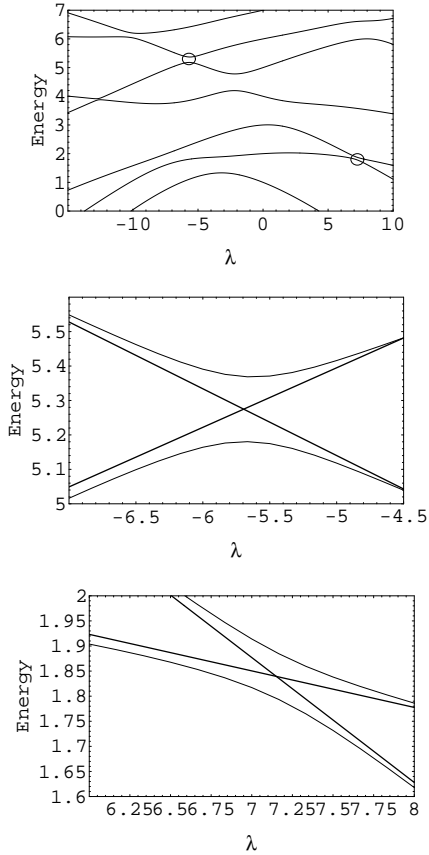}}
\vglue 0.15cm
\caption{
Two-dimensional approximations of the EP associated with level repulsions.
The top drawing displays a section of a ten-dimensional problem. The drawings
in the middle and at the bottom are blow-ups of the encircled areas of the top.
The straight lines are explained in the text. The distinction between the exact
and the effective two-dimensional problem is within the line thickness for 
the curved lines in the middle and bottom drawing.
}
\label{fig1}
\end{figure}

To summarise: an EP is locally equivalent to a two-dimensional problem. Knowing
all parameters of the effective two-dimensional problem we know, from
Eqs.(\ref{ep}) and (\ref{pun}), the specific superposition of the wave
function at the EP in terms of those wave functions which coalesce at the EP.
We find the relation
\begin{eqnarray}
|\psi _{{\rm EP}}\rangle =+i|\chi _{\nu }\rangle &+&|\chi _{\nu +1}\rangle 
\\
{\rm for} \quad \lambda _c^+&=&-{\epsilon_{\nu }-\epsilon_{\nu +1}\over
\omega _{\nu }-\omega _{\nu +1}}e^{+2i\phi },\nonumber  \\
|\psi _{{\rm EP}}\rangle =-i|\chi _{\nu }\rangle &+&|\chi _{\nu +1}\rangle 
\\
{\rm for} \quad \lambda _c^-&=&-{\epsilon_{\nu }-\epsilon_{\nu +1}\over
\omega _{\nu }-\omega _{\nu +1}}e^{-2i\phi }. \nonumber 
\end{eqnarray}
In a higher dimensional problem the quantities $\epsilon _j,\omega _j$ 
and $\phi $ are effective quantities as defined above.

In an experimental situation like a microwave resonator, the phase factor $+i$
means that the time dependent wave function $|\chi _{\nu }\rangle $ has 
a leading phase of a quarter of a full period with respect to 
$|\chi _{\nu +1}\rangle $. For the phase $-i$ the wave is lagging by the 
same amount. This should be detectable \cite{graf}.
In the particular case, where the two wave functions can be associated with
two independent linear polarisations, the wave function at the EP would then
be an elliptic or circular wave with a definite chirality.
We note that a similar observation has been made in \cite{shuv} for the 
treatment of damped acoustic waves in a solid medium. If 
the two wave functions can be associated with different parities, the 
superposition again has a definite chirality. 

We conclude that a definite chirality is associated 
with each EP. In a high dimensional problem one expects a random occurrence
of a particular chiral behaviour just as the random occurrence of the
associated level repulsions. Note that, in an experiment, only those EP are
accessible which have a negative imaginary part of the eigenenergy. 
Depending on the effective values of the $\epsilon _j,\omega _j$ and 
$\phi $, these points may lie in the upper or lower $\lambda $-plane.

{\bf Ackowledgement} WDH grately enjoyed the warm hospitality of the theory 
group at the Max-Planck-Institute for Nuclear Physics at Heidelberg. Both
authors ackowledge stimulating discussions with their experimantal colleagues
at the TU Darmstadt.

\end{document}